\documentclass[namedreferences]{SolarPhysics}
\usepackage[optionalrh]{spr-sola-addons} 
\usepackage{graphicx}        
\usepackage{color}           
\usepackage{url}             

\begin{document}

\begin{article}

\begin{opening}

\title{Physical Properties of Wave Motion in Inclined Magnetic Fields
Within Sunspot Penumbrae}

\author{H.~\surname{Schunker}$^{1,2}$\sep
        D.C.~\surname{Braun}$^{3}$\sep
        C.~\surname{Lindsey}$^{3}$\sep   
        P.S.~\surname{Cally}$^{2}$\sep   
       }
\runningauthor{Schunker et al.}
\runningtitle{Surface Velocities in Magnetic Fields}

   \institute{$^{1}$ Max Planck Institute for Solar System Research, Max-Planck Strasse 2, Katlenburg-Lindau 37197, Germany
                     email: \url{schunker@mps.mpg.de} \\ 
              $^{2}$ Monash University, Melbourne, VIC 3800, Australia 
                    \\
                    $^{3}$ NorthWest Research Associates, Inc., Colorado Research Associates Division, 3380 Mitchell Lane, Boulder, CO 80301 
             }

\begin{abstract}
At the surface of the Sun acoustic waves appear to be affected by the presence of strong magnetic fields in active regions. We explore the possibility that the inclined magnetic field in sunspot penumbrae may convert primarily  vertically propagating
acoustic waves into elliptical motion.  We use helioseismic holography to measure the modulus and phase of the correlation between  incoming acoustic waves and the local surface motion within two sunspots. These correlations are modeled assuming the surface motion is elliptical, and we explore the properties of the 
elliptical motion on the magnetic field inclination.  We also demonstrate that the phase shift of the outward  propagating waves is
opposite to the phase shift of the inward propagating waves in stronger, more
vertical fields, but similar to the inward phase shifts in weaker, more inclined
fields.
\end{abstract}
\end{opening}

\section{Introduction}
     \label{introduction}
     Helioseismology uses the observed solar surface acoustic wavefield to construct 
images of the subsurface structure of the Sun. Of particular interest
has been the three-dimensional (3D) modeling of time-distance 
\cite{DJHP93} observations
of travel-time shifts to deduce the subsurface structure of active regions 
\cite{KDS00,ZK03}.  Assuming the travel-time shifts are due to perturbations in
the sound speed below the spot, a general consensus in the models has
emerged consistent with sound-speed reductions (relative to surrounding quiet
Sun) near the surface ($\lesssim 4$ Mm) and enhancements up to 15 Mm below 
sunspots \cite{KDS00,CBK06}.  Using ring diagram analysis 
\inlinecite{BAB04}  also find a lower sound speed immediately below the 
surface and an increase in the sound speed below 7 Mm.

A comparison between Fourier-Hankel analysis and time-distance results 
by \inlinecite{B97} first prompted caution in the interpretation of acoustic 
oscillation signals within sunspots. 
The influences of strong surface magnetic fields have not been explicitly
included in most helioseismic models of active regions. 
 \citeauthor{LB05ii} (2005a) and \citeauthor{LB05i} (2005b) have shown
that helioseismic phase shifts observed with helioseismic
holography vanish below a depth of about 5 Mm, 
when a surface (``showerglass'') phase shift
based on photospheric magnetic flux density, is removed from  
the data.

Other evidence supports the possibility of strong near-surface
contributions to the helioseismic phase (or travel-time) shifts.
These include the possible contamination of surface perturbations
into the 3D inversions (e.g. \opencite{K06}; 
\opencite{CR07}). It has also been shown that
the reduction of $p$-mode amplitudes in magnetic regions
can cause travel-time shifts \cite{RBDTZ06}. The
suppression of sources of wave excitation within sunspots
can also produce measurable shifts \cite{HCRB07}.

\inlinecite{SBCL05} and \inlinecite{SBC07} have found that
phase shifts obtained from seismic holography in sunspot
penumbrae vary with the line-of-sight angle (from vertical) 
as projected into the plane containing the magnetic field and the 
vertical direction.  A similar effect has also been noted by 
\inlinecite{ZK06} with time distance measurements.
\inlinecite{SBC07} find that the  effect is dependent upon the strength and/or 
inclination of the magnetic field. 
In the penumbra the magnetic field strength 
decreases as the magnetic field angle from vertical increases, hence the two 
properties of the magnetic field cannot be extricated. 
The phase variation with line-of-sight 
viewing angle is demonstrated most substantially at frequencies 
around 5 mHz with a 
strong, almost vertical magnetic field close to the umbra.  
\inlinecite{SBC07} also find that the total variation across
all lines-of-sight increases with temporal frequency,
particularly in the stronger fields in the penumbrae.

Mode conversion of the acoustic waves in the near surface has been explored as 
the physical cause of the observed absorption of acoustic waves by sunspots.  A 
fast acoustic wave, propagating towards the surface from the interior, 
encounters the depth at which the Alfv\'en speed is equal to the sound speed 
($a\approx c$) which is typically close to the surface in a  sunspot. Under 
these conditions it is able to transmit to a slow acoustic mode and convert to 
a fast magnetic mode \cite{C05}. \inlinecite{CC03} explore the mode 
conversion in two-dimensions with a uniform \emph{inclined} magnetic field 
relevant to sunspot penumbrae. The inclination of the magnetic field is found 
to have a significant dependence on the likelihood of conversion  and fits 
extremely well with the analysis of \inlinecite{B95} \cite{CCB03}. Further 
work \cite{C05,SC06} using ray theory has since established that it is the 
angle between the acoustic wave path and the magnetic field (the `attack 
angle') which is the crucial factor inducing conversion. With a wide attack 
angle at the $a\approx c$ level there is maximum conversion from a fast 
acoustic to a fast magnetic mode. A fine attack angle encourages  transmission  to a slow acoustic mode, which is guided `up' the magnetic field lines to 
observational heights in the atmosphere. A consequence of mode conversion
may be the observational signature of elliptical motion in regions
of inclined magnetic field. 

The aim of this paper is to model the observations of two
sunspots previously analyzed by \inlinecite{SBC07} in order to determine the properties of velocity ellipses consistent with the data.
These models are based on a least-squares-fit of the phase and 
modulus information of the local ingression control correlation. 
We explore the properties of these ellipses, observed
with waves at different temporal frequencies, as functions of
the magnetic field inclination angle. 
We also examine the variation with line-of-sight angle
of the phase shifts in the outgoing waves, using the local egression 
control correlation, to assess the relation of the phase shift
variations between incoming and outgoing waves.

In the following sections we describe the data (Section \ref{obs}), give an 
outline of the helioseismic holography technique used (Section 3), describe the 
results (Section 4) and discuss our results in the
context of mode conversion  (Section 5).

 \section{Observations} 
      \label{obs}      

As this is a continuation of studies by \inlinecite{SBCL05} and 
\inlinecite{SBC07} we use the same data, however, we provide a short 
description here to maintain coherence. The \emph{Michelson Doppler Imager} 
(MDI) aboard the \emph{Solar and Heliospheric Observatory} (SOHO) \cite{MDI95} 
provides the solar surface Doppler velocity information. The Dopplergrams are 
full disk, have a 60 second cadence and  a resolution of  $\approx 1.4$ Mm per 
pixel. These full disk Dopplergrams are Postel projected and a $512\times 512$ 
pixel extract is taken centered on the active region. We analyze two sunspots: 
the first in AR9026  observed over 10 days from 3rd - 12th June 2000 with a 
Carrington longitude (L0) of $75^\circ$ and latitude (B0) of $20^\circ$ and 
penumbral boundaries defined by inner and outer radii of 7 Mm and 16 Mm 
respectively. The second is a sunspot in AR9057 observed over 9 days from 24th 
June - 2nd July 2000 with $L0=158^\circ$ and $B0=13^\circ$, inner and outer
penumbral boundaries given by 6 Mm and 13 Mm.  
The penumbral boundaries are 
defined to be between 50\% and 85\% of the nearby quiet-Sun 
continuum intensity.
The sunspots in these active regions were chosen 
based on the existence of continual MDI observation as they traversed the solar 
disk, and for their relatively simple magnetic structure and evolution. 

Magnetograms from the \emph{Imaging Vector Magnetograph} (IVM) at the 
University of Hawaii Mees Solar Observatory \cite{IVM96}  provide the 
orientation and strength of the surface magnetic field in the sunspots chosen 
in AR9026 and AR9057. The IVM observations are made over a 28 minute interval: 
for AR9026  starting at 18:29 Universal Time (UT) on 5th June 2000 and  for 
AR9057 starting at 16:19 UT  on 28th June 2000. 
The IVM data reveals an azimuthally spreading magnetic field configuration 
for both sunspots although they are not entirely symmetric.  It is 
assumed, supported largely by available line-of-sight magnetograms, that there 
is no significant evolution of the magnetic field in the sunspots during the 
time of observation. Therefore, using only one vector magnetogram for the 
duration of the observation is reasonable. Rotation and scaling are applied to 
align the IVM data to the line-of-sight MDI magnetograms.

\begin{figure}[ht]
\begin{center}
\includegraphics[width=0.8\textwidth]{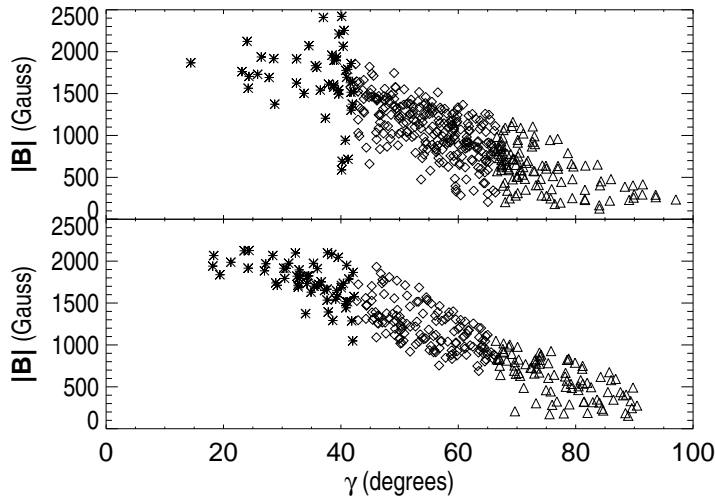}
\vspace{0cm}
\caption{Magnetic field strength, $| B |$, plotted against inclination from 
vertical, $\gamma$, determined from IVM vector magnetograms. AR9026 (5th June 
2000) (bottom) and AR9057 (28th June 2000) (top). The different symbols divide 
the penumbra into roughly equal regions of inclination: $\gamma < 42^\circ$ 
(asterisk); $42^\circ < \gamma < 66^\circ$ (diamonds); $\gamma > 66^\circ$ 
(triangles). This corresponds to average magnetic field strengths of 1700 G, 
1000 G and 600 G for AR9057 and for AR9026 1900 G, 1400 G  and 600 G. In 
general, the progression from the upper left portion of the distribution to the 
lower right portion represents increasing distance from the centre of the spot.}
\label{g_vs_b}
\end{center}
\end{figure}

Figure~\ref{g_vs_b} shows a strong correlation between the 
magnetic field inclination (from vertical, $\gamma$) and 
field strength (the strong field is almost vertical whereas highly 
inclined field is relatively weak). In 
this paper we use the inclination $\gamma$ as the primary variable,
dividing the penumbrae into three regions defined by the values
of $\gamma$ (Fig~~\ref{g_vs_b}).
It is understood in this analysis that the magnetic field strength is 
implicitly correlated with the inclination through Figure~\ref{g_vs_b}, 
and that independent dependencies of observables with field strength 
and inclination are not extracted. The 
dependence of the phase shifts on the line-of-sight viewing angle of the 
magnetic field is facilitated by knowing the full vector magnetic field.

\section{Helioseismic Holography}

Helioseismic holography \cite{LB00,BL00} is the phase coherent imaging of the 
solar subsurface based on photospheric acoustic oscillations. The ingression is 
an assessment of the observed wavefield, $\psi(\mathbf{r}',t)$, converging to a 
selected focal point, $(\mathbf{r},z,t)$, and the egression is the time reverse 
- an assessment of waves diverging from that point. In this case we calculate 
the quantities at the surface, $z=0$. In practice, the observed wavefield used 
for the calculation is usually an annulus surrounding the chosen focal point. 
The pupil used here is identical to that described by \inlinecite{SBCL05} and is 
constructed for the  calculations with inner radius $a=20.7$ Mm  and outer 
radius $b= 43.5$ Mm, designed to be large enough that when the focal point is 
within the penumbra the area covered by the annulus does not include large 
areas of strong magnetic field.  At a frequency of 5 mHz this selects $p$-modes 
with spherical harmonic degree and radial degree between $\ell \approx 450$ and 
$\ell \approx 700$.  The ingression at the surface is given by,
\begin{equation}
H_{-}(\mathbf{r},0,t)=\int_{a < |\mathbf{r}-\mathbf{r'}|<b} d^2 \mathbf{r}' 
G_{-}( |\mathbf{r} - \mathbf{r}' |, 0, t - t') \psi(\mathbf{r}',t),
\label{ingress}
\end{equation}
$G_{-}$ is the ingression Green's function \cite{LB00}. The egression, $H_{+}$ 
is simply the time reverse of  \ref{ingress}, i.e. $t' - t$. 

The ingression (\ref{ingress}) is correlated with the observed wavefield in the 
space-frequency domain,
\begin{equation}
C_{-}(\mathbf{r},\nu) = \langle \hat{H}_{-} 
(\mathbf{r},\nu)\hat{\psi}^{*}(\mathbf{r},\nu ) \rangle_{\Delta \nu} = 
|C_{-}|  \rm{e}^{-i \ \delta \phi_{-}},
\label{correl}
\end{equation}
where $\hat{H}_{-}(\mathbf{r},\nu)$ and $\hat{\psi}(\mathbf{r},\nu )$ are
the temporal Fourier transforms of $H_{-}$ and $\psi$ respectively,
and we have dropped the dependence on depth $z$.
The correlation has a modulus $|C_{-}|$ and 
a phase, $\delta \phi_{-}$.   
In the frequency spectrum, with the pupil covering mostly quiet Sun, the local 
ingression control correlation simply characterizes how the local magnetic 
photosphere responds acoustically to upcoming waves, prescribed by $H_{-}$, 
originating within the pupil.

In \inlinecite{SBCL05} and \inlinecite{SBC07} the phase shift caused by the 
surface perturbations to the incident wave is calculated at various 
frequencies. Here we extend that research and monitor the variation of the 
correlation amplitude within the sunspot penumbral region, which is then 
combined with the phase information to form estimates of the surface velocity 
ellipse. We then go on to examine the phase of the local egression control 
correlation, $\delta \phi_{+} = \rm{Arg} [ C_{+}(\mathbf{r},\nu)]$, with the 
line-of-sight angle in the penumbra of sunspots in AR9057 and AR9026.

\section{Elliptical Representation of Surface Velocities}

For better statistics multiple days of observation of each sunspot are 
combined and a least-squares-fit of the observations allows an estimation of 
the velocity ellipse. To calculate the velocity ellipse  
the ingression correlation (eqn. \ref{correl}) 
is used as a proxy for the local velocity. The modulus and phase
of the correlation vary with respect 
to the line-of-sight angle and these variations are modeled to construct 
an ellipse representing the 
surface velocity associated with a particular magnetic field element. 

A smeared ingression `flat field' takes out undesired contributions to the 
correlation modulus due to the temporal or spatial variations of the
ingression, due to the presence of magnetic
regions in the pupil. This is achieved by dividing $|C_{-}|$ by the 
Gaussian smear of the root-mean-square of the ingression to get 
$|C_{-}|_{flat}$. The resulting correlations are normalized so that 
the ingression correlation modulus in the nearby quiet Sun 
has a value of $\cos(\zeta)$, where $\zeta$ is the heliocentric
angle of the quiet region from disk center. A dependence of the
quiet Sun (root-mean-squared) amplitude with  $\cos(\zeta)$ is
expected for predominantly vertically oscillating $p$-modes. 
The normalization is achieved by dividing $|C_{-}|_{flat}$ 
by the quiet Sun average divided by the cosine of the heliocentric angle, i.e. 
\begin{equation}
|C_{-}|_{norm}=|C_{-}|_{flat} \frac{\cos(\zeta)}{\langle |C_{-}| 
\rangle_{QS}}. 
\end{equation}
The normalization corrects for variations in the modulus due to 
duty cycle variations, foreshortening, and other effects which 
cause undesired variations in the modulus from day to day. 
These procedures assume that these 
factors have the same relative effect on the (desired) correlations in the 
penumbra as they do to the quiet Sun. It is difficult to assess the validity 
of this, and so it is used as a reasonable working assumption subject to some 
caution.

We use the same angle $\theta_p$ as defined in \inlinecite{SBCL05}, which is 
the angle between the projection of the line-of-sight vector onto the plane 
containing the magnetic field vector and the radial vector. 
It is a measure of the angle that the magnetic field is being viewed from. We 
now  simply refer to $|C_{-}|_{norm}$ as $|C_{-}|$ and see how it varies with 
$\theta_p$.

\begin{figure}[h]
\begin{center}
\includegraphics[width=0.7\textwidth]{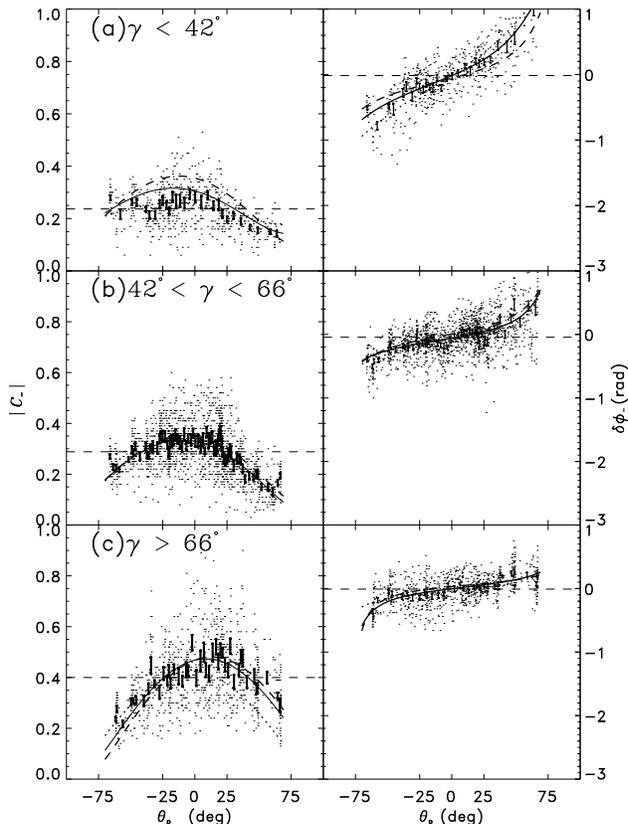} \hspace{1.5cm}
\caption{The modulus of the correlation, $|C_{-}|$  (left column), and the 
phase, $\delta \phi_{-}$, in the penumbra of AR9026 at 3 mHz for all days of 
observation are plotted against projected angle $\theta_p$ for different values 
of magnetic field inclinations as indicated. The top panel (a) shows $\gamma < 
42^\circ$, where the mean field strength is $\langle \mathbf{B} \rangle = 1900$ 
G, the middle panel (b) shows $42^\circ < \gamma < 66^\circ$, where $\langle 
\mathbf{B} \rangle = 1400$ G, and the bottom panel (c) shows  $\gamma > 
66^\circ$, where $\langle \mathbf{B} \rangle = 600$ G. The horizontal dashed 
lines indicate the mean value of $|C_{-}|$ for each panel. The error bars 
indicate the standard deviation of the mean over bins of 20 measurements in 
$\theta_p$. The solid line of fit is a fit for all the displayed data; the 
dotted line is a fit for the data from 3rd - 7th June 2000; the dashed line is 
a fit for data from 8th - 12th June 2000.  }
\label{3mhz}
\end{center}
\end{figure}

\begin{figure}[ht]
\begin{center}
\includegraphics[width=0.7\textwidth]{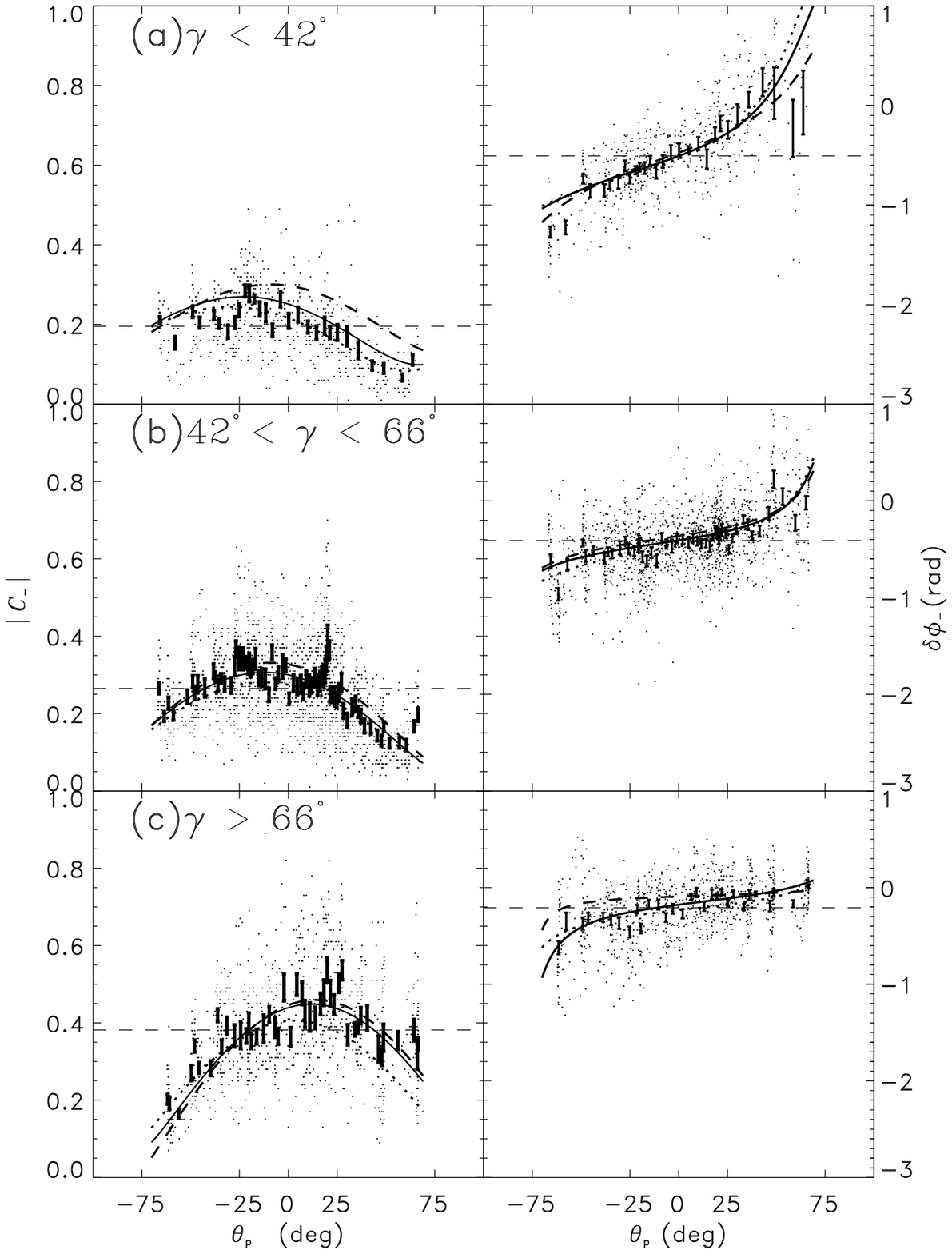}
\caption{Same as Figure~\ref{3mhz} except at 4 mHz.}
\label{4mhz}
\end{center}
\end{figure}

\begin{figure}[ht]
\begin{center}
\includegraphics[width=0.7\textwidth]{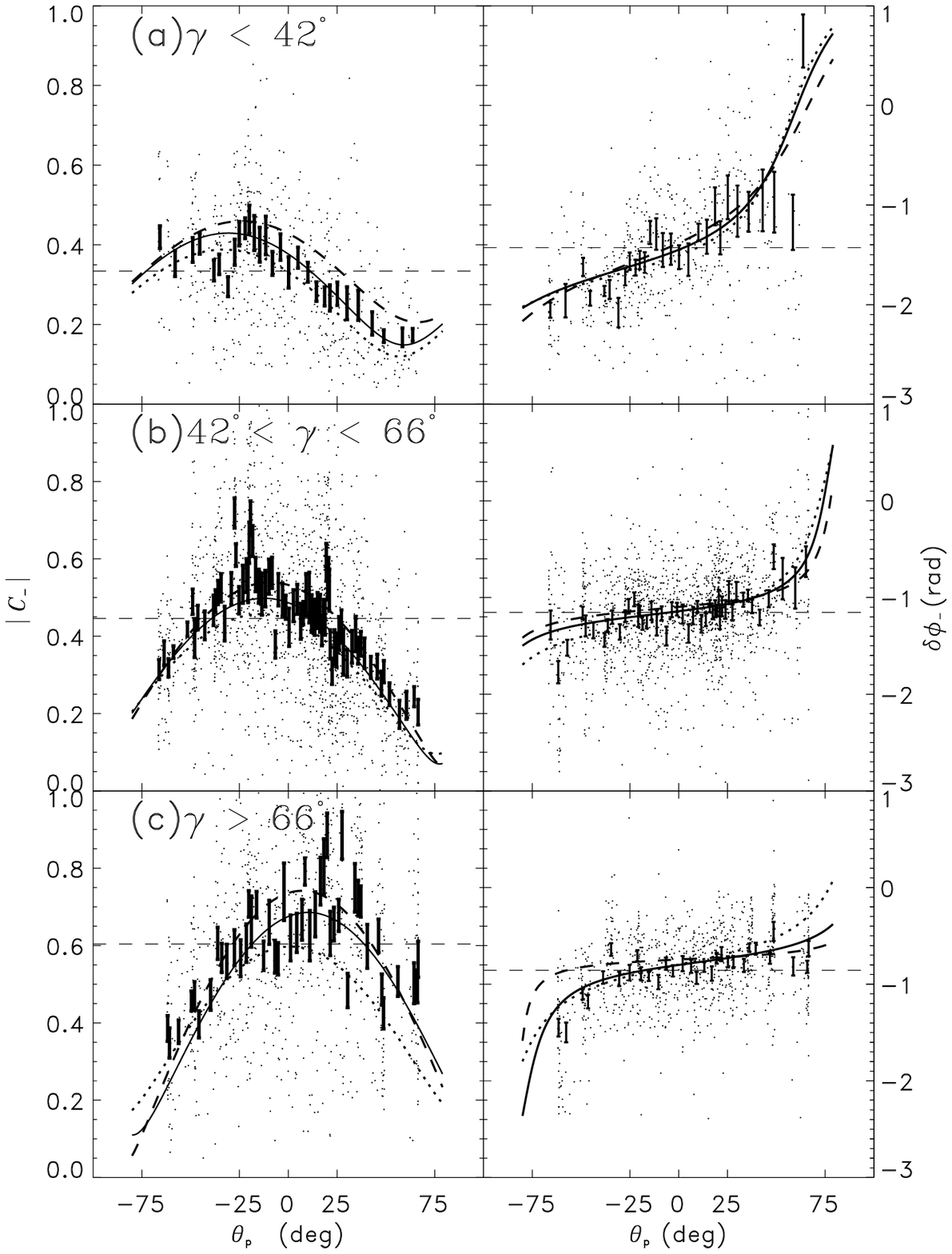}
\caption{Same as Figure~\ref{3mhz} except at 5 mHz.}
\label{5mhz}
\end{center}
\end{figure}

\begin{figure}[ht]
\begin{center}
\vspace{1cm}
\includegraphics[width=0.7\textwidth]{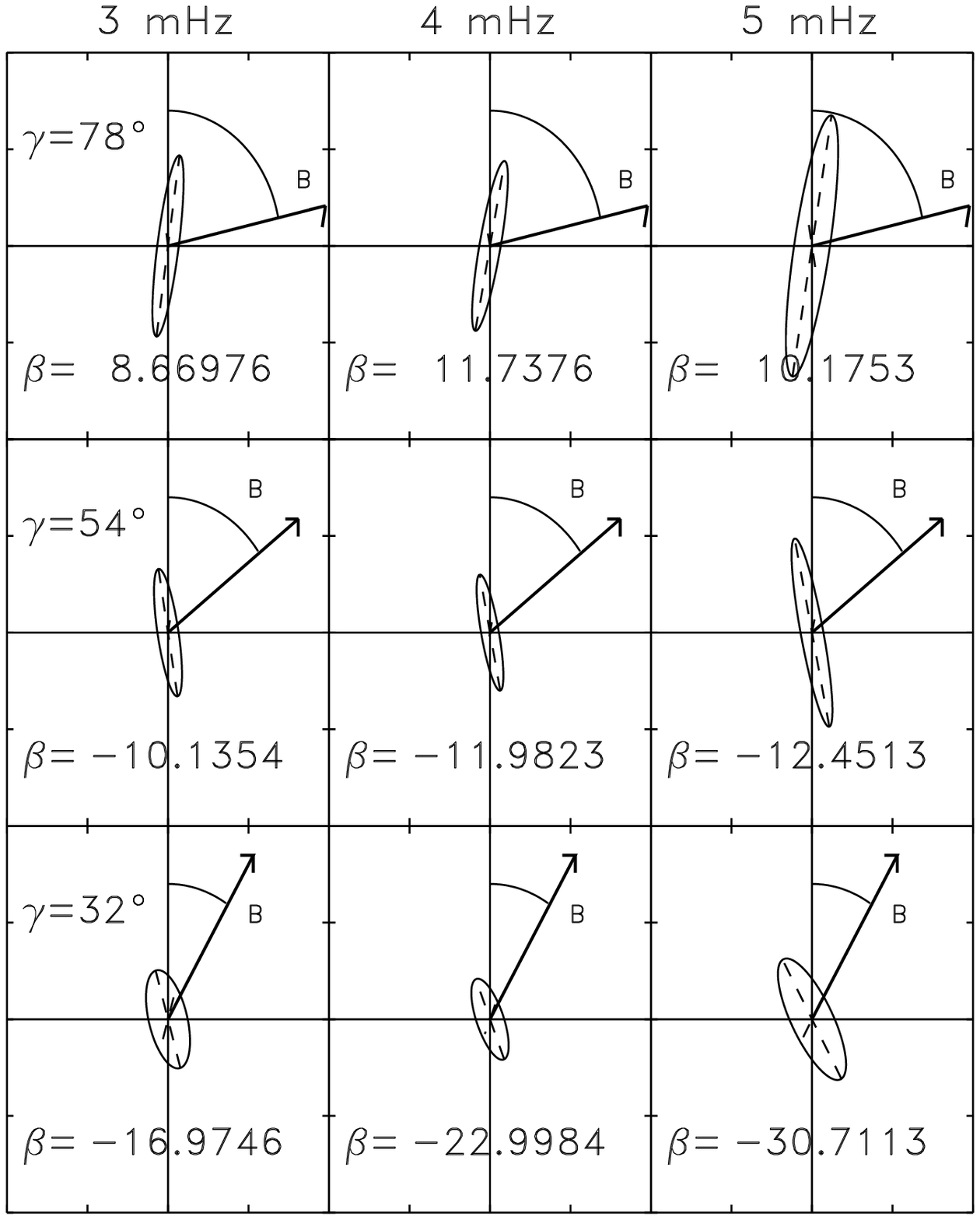}
\caption{Least-squares-fit surface velocity ellipses as given by the phase and 
amplitude of the local ingression control correlation in the penumbra of 
AR9026. The top row is when $\gamma > 66^\circ$, the middle row when $42^\circ 
< \gamma < 66^\circ$ and the bottom row is when $\gamma < 42^\circ$. The 
$\gamma$ listed in the plot is the angle that the magnetic field vector is 
drawn at. $\beta$ is the inclination angle of the semi-major axis of the 
velocity ellipse. The left column is at frequencies of 3 mHz, the middle column 
at 4 mHz and the right column at 5 mHz.}
\label{ellipse9026}
\end{center}
\end{figure}

\begin{figure}[h]
\begin{center}
\includegraphics[width=0.5\textwidth]{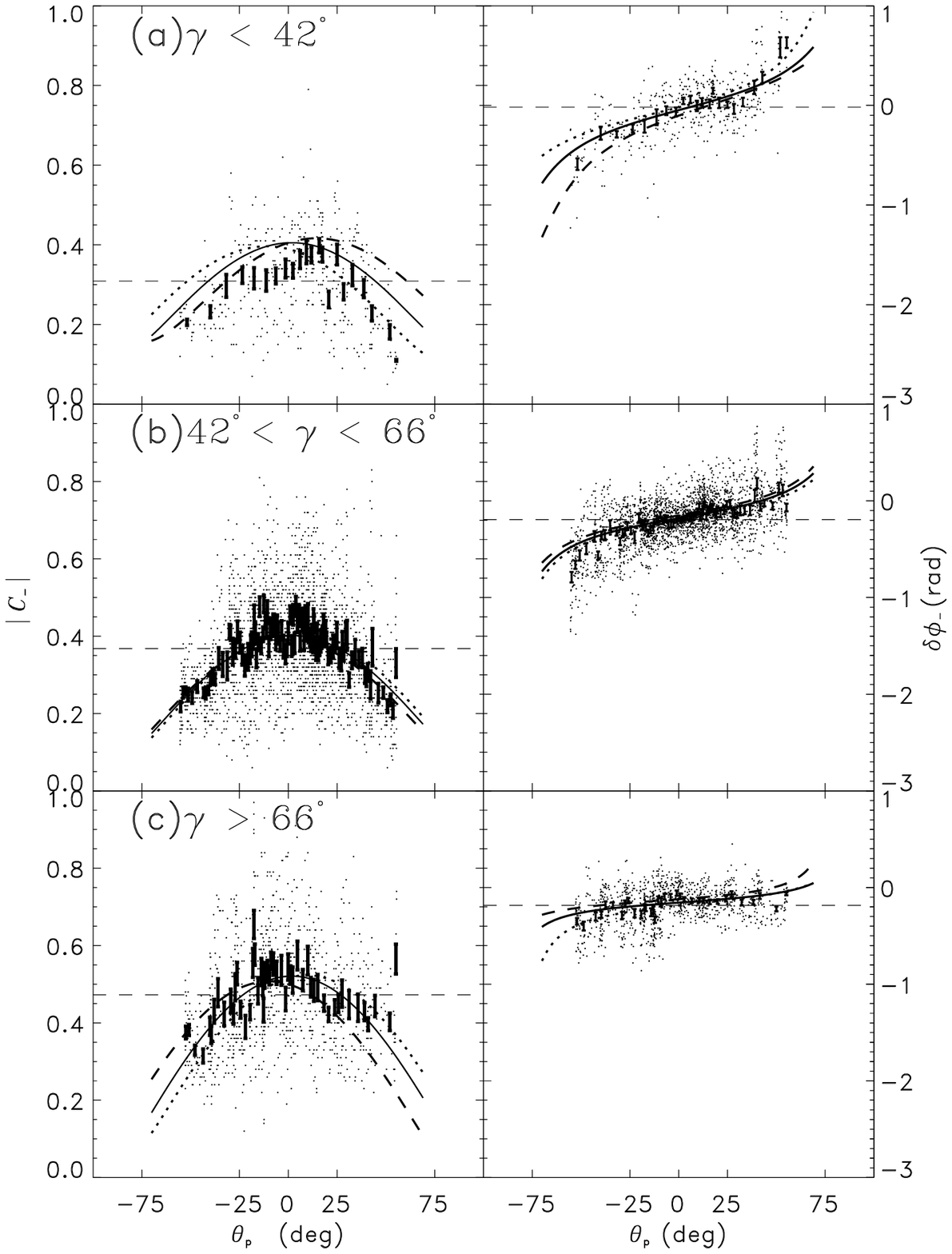}
\caption{The modulus of the correlation, $|C_{-}|$ (left column) and phase, 
$\delta \phi_{-}$ (right column), in the penumbra of AR9057 at  3 mHz for all 
days of observation are plotted against projected angle $\theta_p$ for 
different values of magnetic field inclinations as indicated.  The top panel 
(a) shows $\gamma < 42^\circ$, where the mean field strength is $\langle 
\mathbf{B} \rangle = 1700$ G, the middle panel (b) shows $42^\circ < \gamma < 
66^\circ$, where $\langle \mathbf{B} \rangle = 1000$ G, and the bottom panel 
(c) shows  $\gamma > 66^\circ$, where $\langle \mathbf{B} \rangle = 600$ G. The 
horizontal dashed lines indicate the mean value of $|C_{-}|$ for each panel. 
The error bars indicate the standard deviation of the mean over bins of 20 
measurements in $\theta_p$. The solid line of fit is a fit for all the 
displayed data; the dotted line is a fit for the data from 24th - 28th June 
2000; the dashed line is a fit for data from 29th June - 2nd July 2000.  }
\label{90573mhz}
\end{center}
\end{figure}

\begin{figure}[h]
\begin{center}
\includegraphics[width=0.5\textwidth]{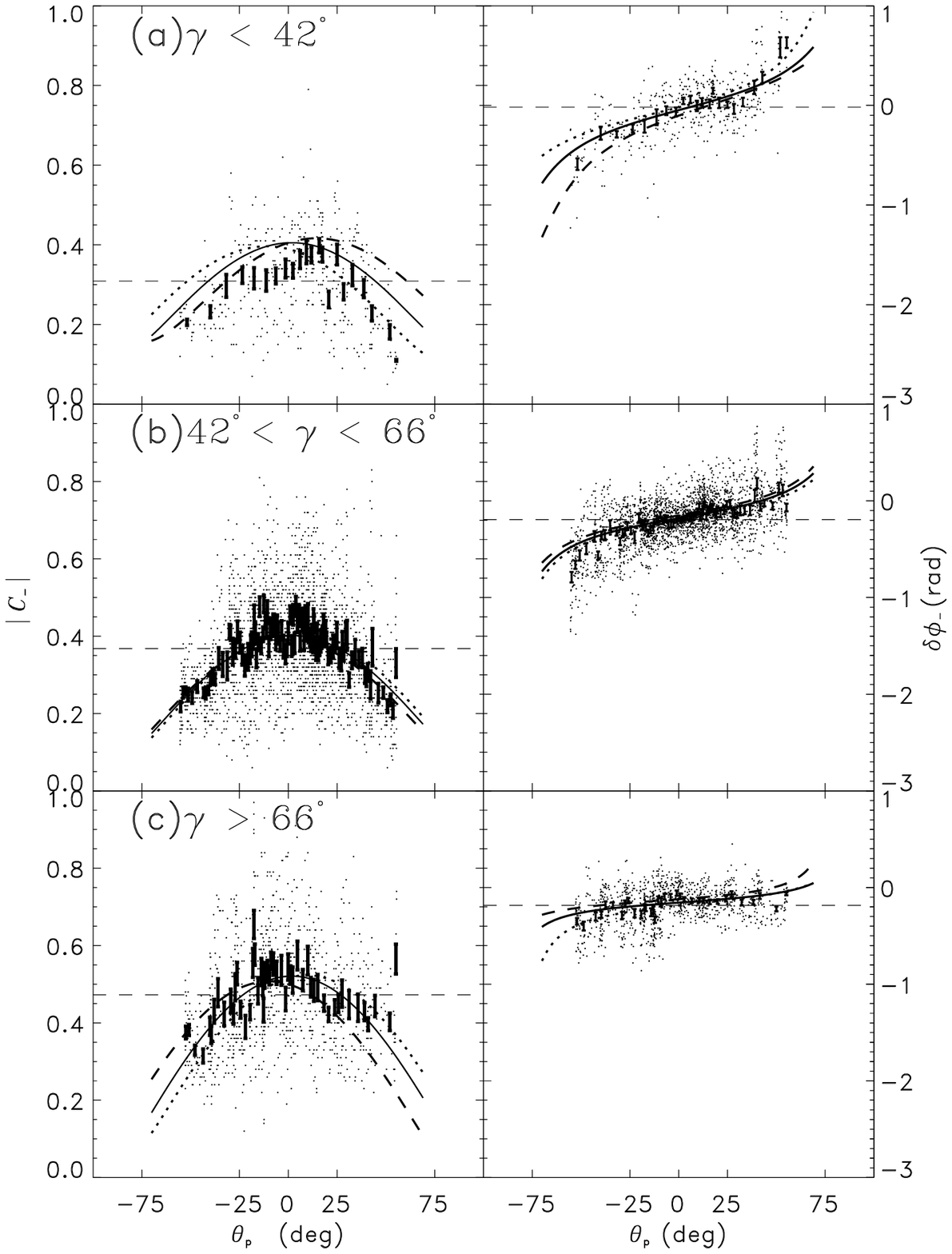}
\caption{Same as for  Figure~\ref{90573mhz} except at 4 mHz. }
\label{90574mhz}
\end{center}
\end{figure}

\begin{figure}[h]
\begin{center}
\includegraphics[width=0.5\textwidth]{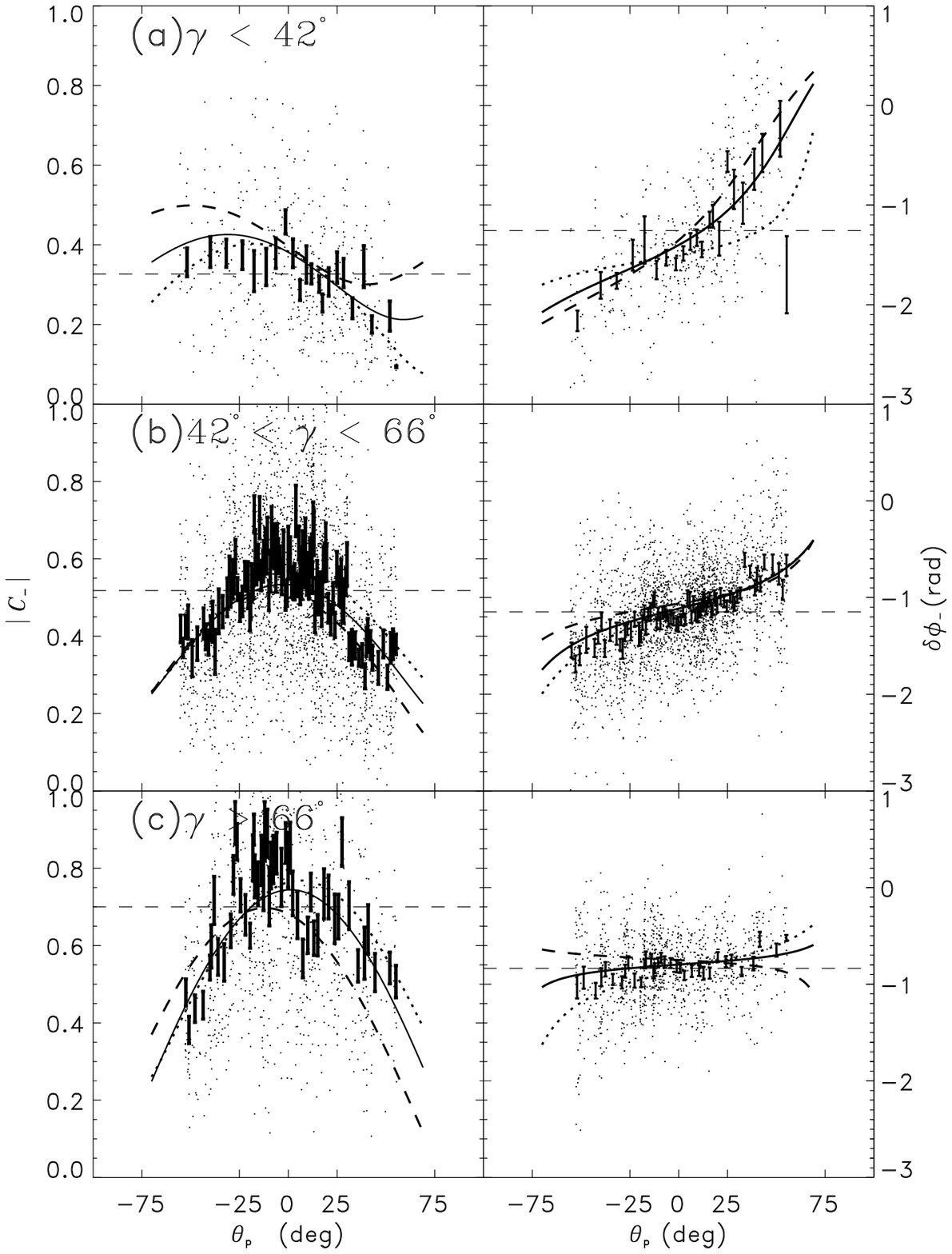}
\caption{Same as for Figure~\ref{90573mhz} except at 5 mHz.}
\label{90575mhz}
\end{center}
\end{figure}

\begin{figure}[h]
\begin{center}
\vspace{1cm}
\includegraphics[width=0.7\textwidth]{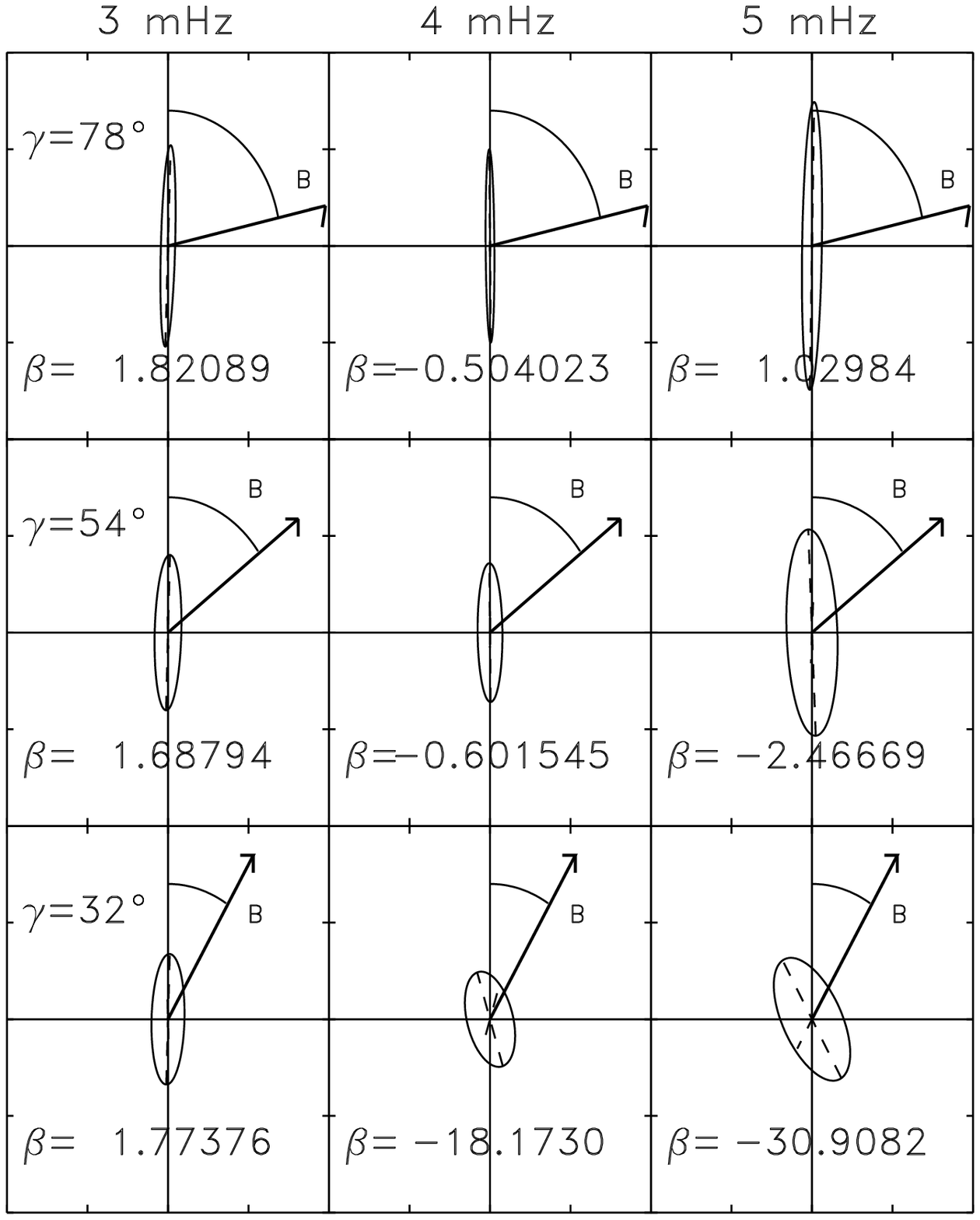}
\caption{Least-squares-fit surface velocity ellipses as given by the phase and 
amplitude of the local ingression control correlation in the penumbra of 
AR9057. The top row is when $\gamma > 66^\circ$, the middle row when $42^\circ 
< \gamma < 66^\circ$ and the bottom row is when $\gamma < 42^\circ$. The 
$\gamma$ listed in the plot is the angle that the magnetic field vector is 
drawn at. $\beta$ is the inclination angle of the semi-major axis of the 
velocity ellipse. The left column is at frequencies of 3 mHz, the middle column 
at 4 mHz and the right column at 5 mHz.}
\label{ellipse9057}
\end{center}
\end{figure}

Figures~\ref{3mhz} to \ref{5mhz} and Figures~\ref{90573mhz} to \ref{90575mhz} 
show the variation of $|C_{-}|$  with $\theta_p$ for AR9026 and AR9057 in the 
left columns at 3, 4 and 5mHz in the same three bins of inclination shown in 
Figure~\ref{g_vs_b}. The right column shows $\delta \phi_{-}$. 
The variations of $\delta \phi_{-}$ with $\theta_p$ have been the subject
of our previous analyses (\opencite{SBCL05}; \opencite{SBC07}). 
In the absence of magnetic effects, we expect
the local oscillatory wave field, as assessed by the ingression correlation,
to be consistent with purely vertical motion. This would predict a
 dependence of $|C_{-}|$ on $\cos(\theta_p)$ and no dependence  of  $\delta \phi_{-}$ on
$\cos(\theta_p)$. 
Departures from these expectation
are clearly visible in Figs~\ref{3mhz} - \ref{5mhz} and \ref{90573mhz} - 
\ref{90575mhz}. A rough understanding of these results can be
gained by noting that the net variation in the phase shift with
$\theta_p$ is related to the eccentricity of the ellipse, while
the value of $\theta_p$ for maximum $|C_{-}|$ determines the 
orientation of the semi-major axis.
We perform a least-squares-fit to the observed $|C_{-}|$ 
and $\delta \phi_{-}$ to determine the elliptical motion 
consistent with the data. The best-fit ellipses 
are shown in Figures~\ref{ellipse9026} and \ref{ellipse9057}. 
The moduli and phase of $C_-$ as determined from the fits are plotted (as solid
lines) with the data points in Figs 2-4 and 6-8. Separate fits were performed for
independent  5-day subsets of the data (also shown in the Figures by the dotted
and dashed lines). There are no obvious systematic differences in the fits
over time. The orientations and eccentricities are highly consistent across
all frequency bands in both sunspots (Figure~\ref{ellipse9026} 
and Figure~\ref{ellipse9057}). Some systematic differences
between the two spots are evident.
For AR9026, the ellipses are aligned slightly towards
the magnetic field direction at high field inclination 
but `swing' over as the inclination becomes smaller (and the 
magnetic field stronger). For AR9057, the motion is nearly vertical
at low field inclination, but also tilts away from the field
in the stronger, more vertical, fields. 
In both spots, the eccentricity
of the ellipses increases with decreasing field strength or increasing 
inclination.

\subsection{ELLIPSE PARAMETERS}

\begin{figure}[ht]
\begin{center}

\includegraphics[width=0.4\textwidth]{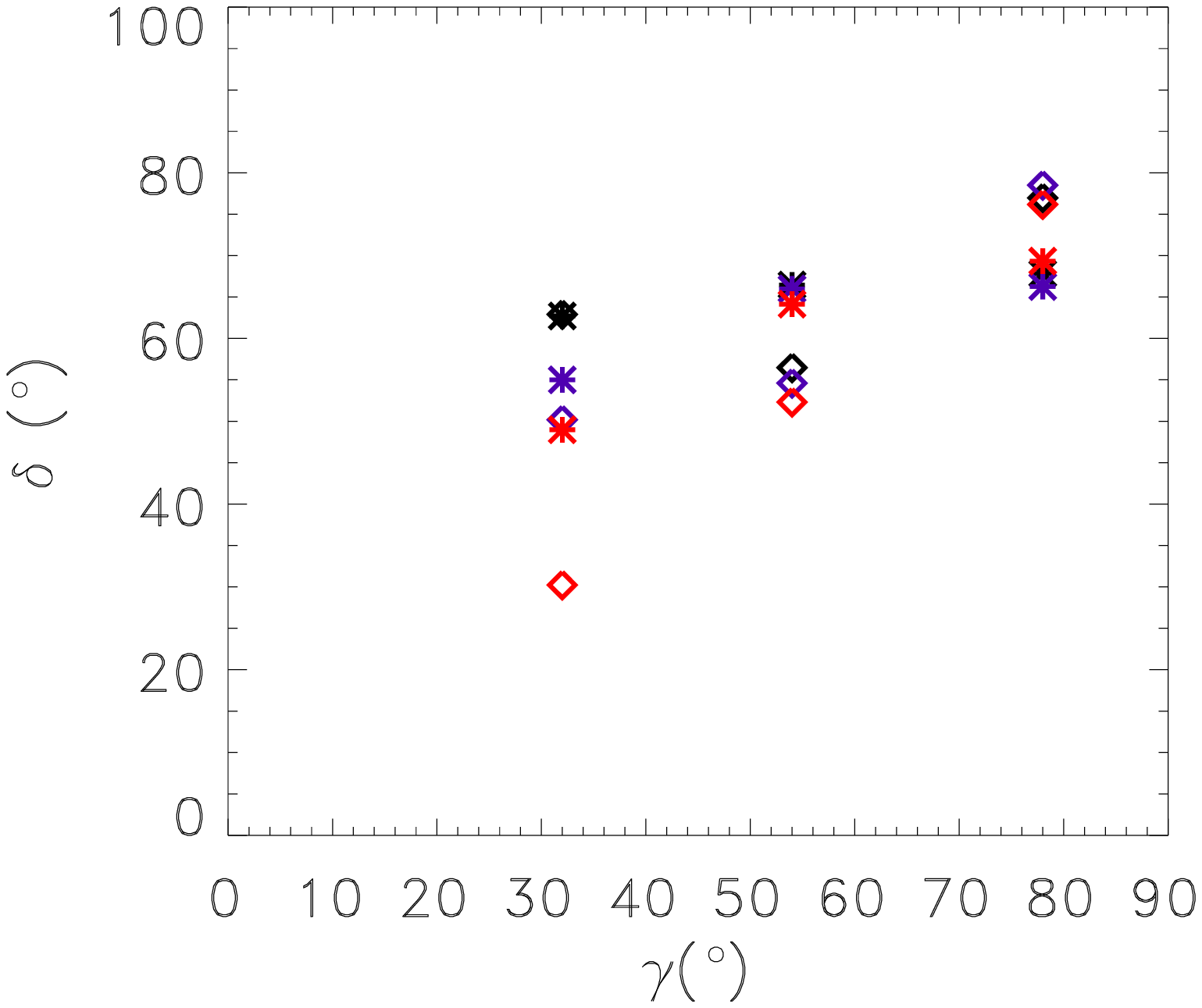}
\includegraphics[width=0.4\textwidth]{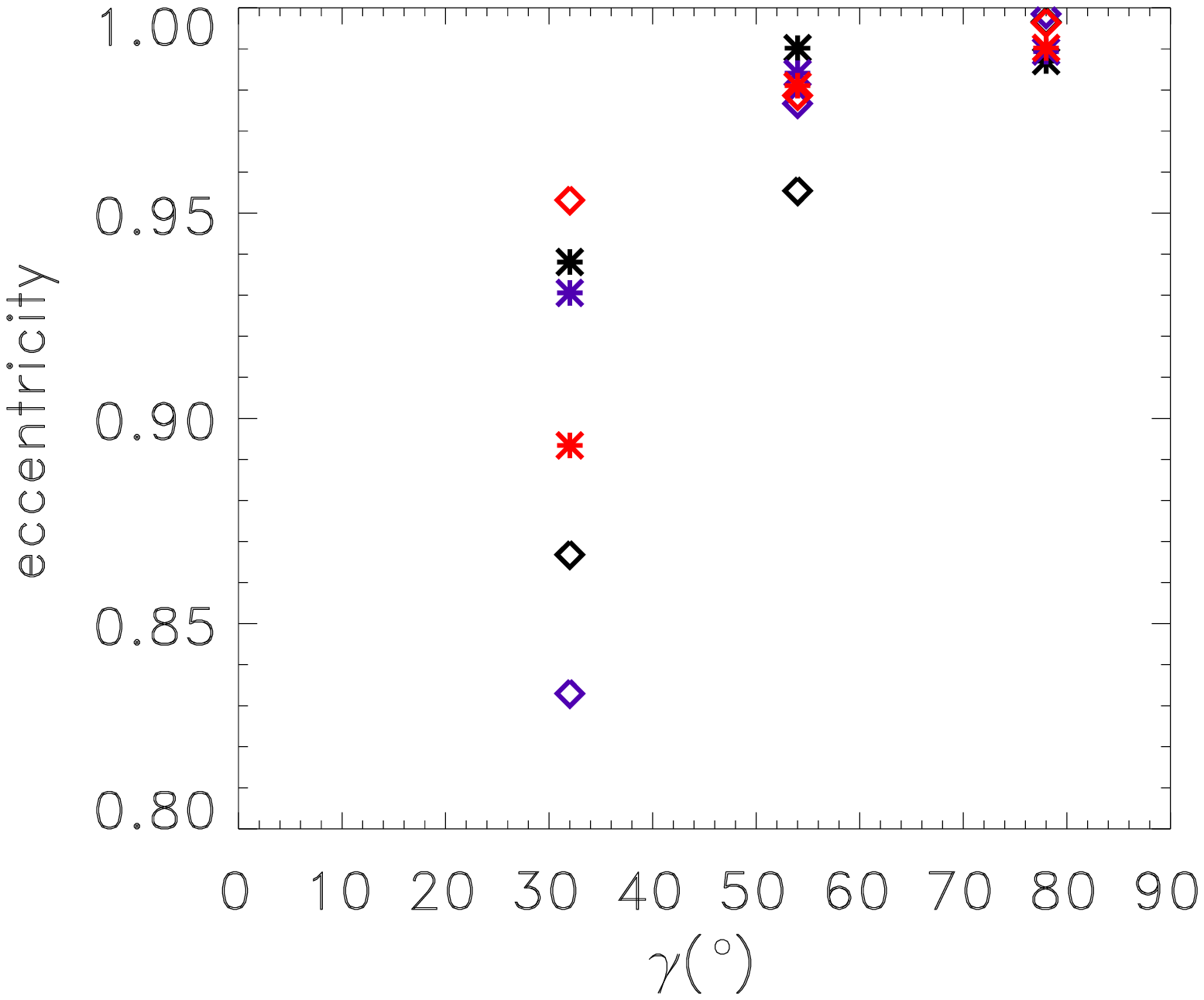}\\
\includegraphics[width=0.4\textwidth]{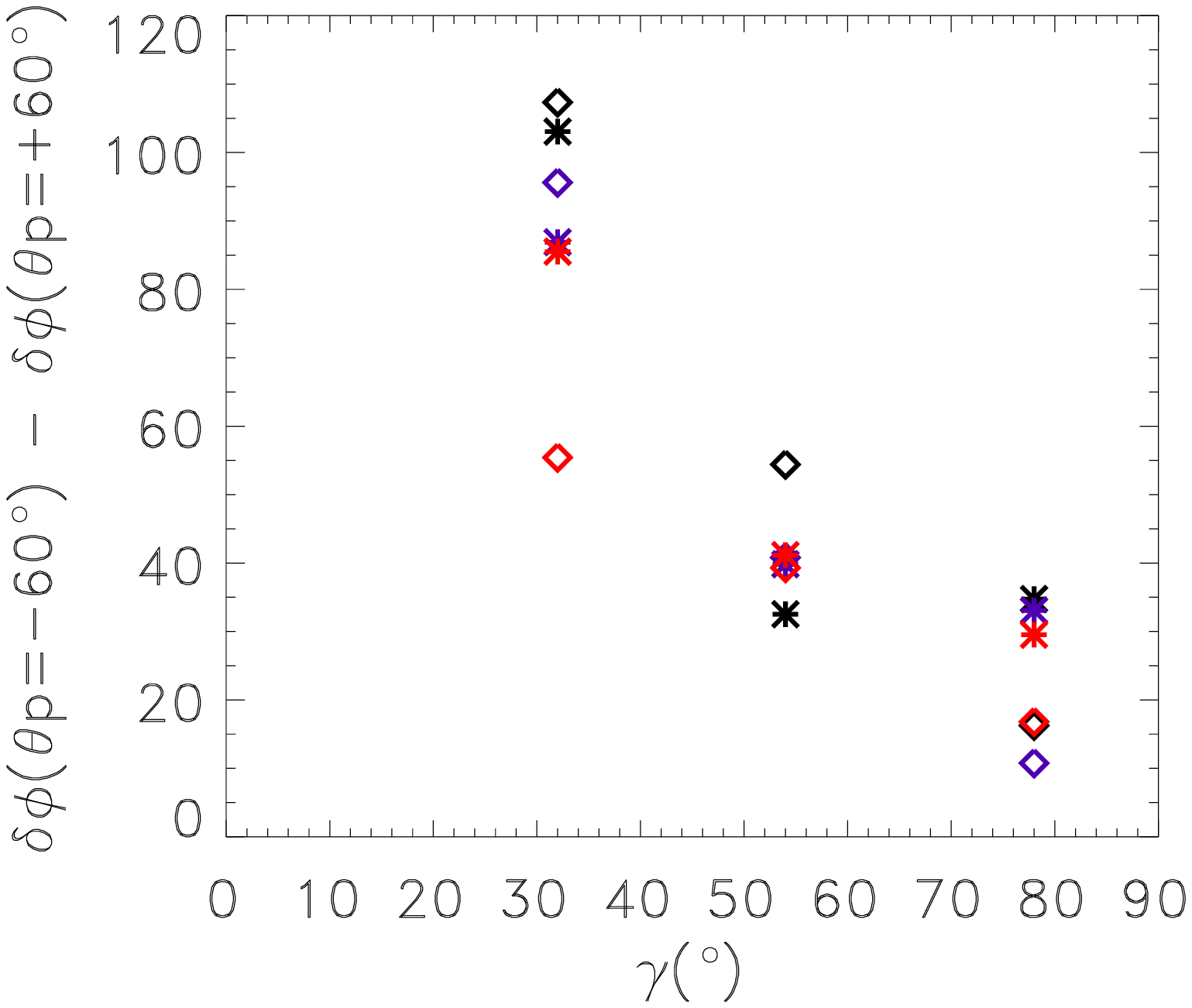}
\includegraphics[width=0.4\textwidth]{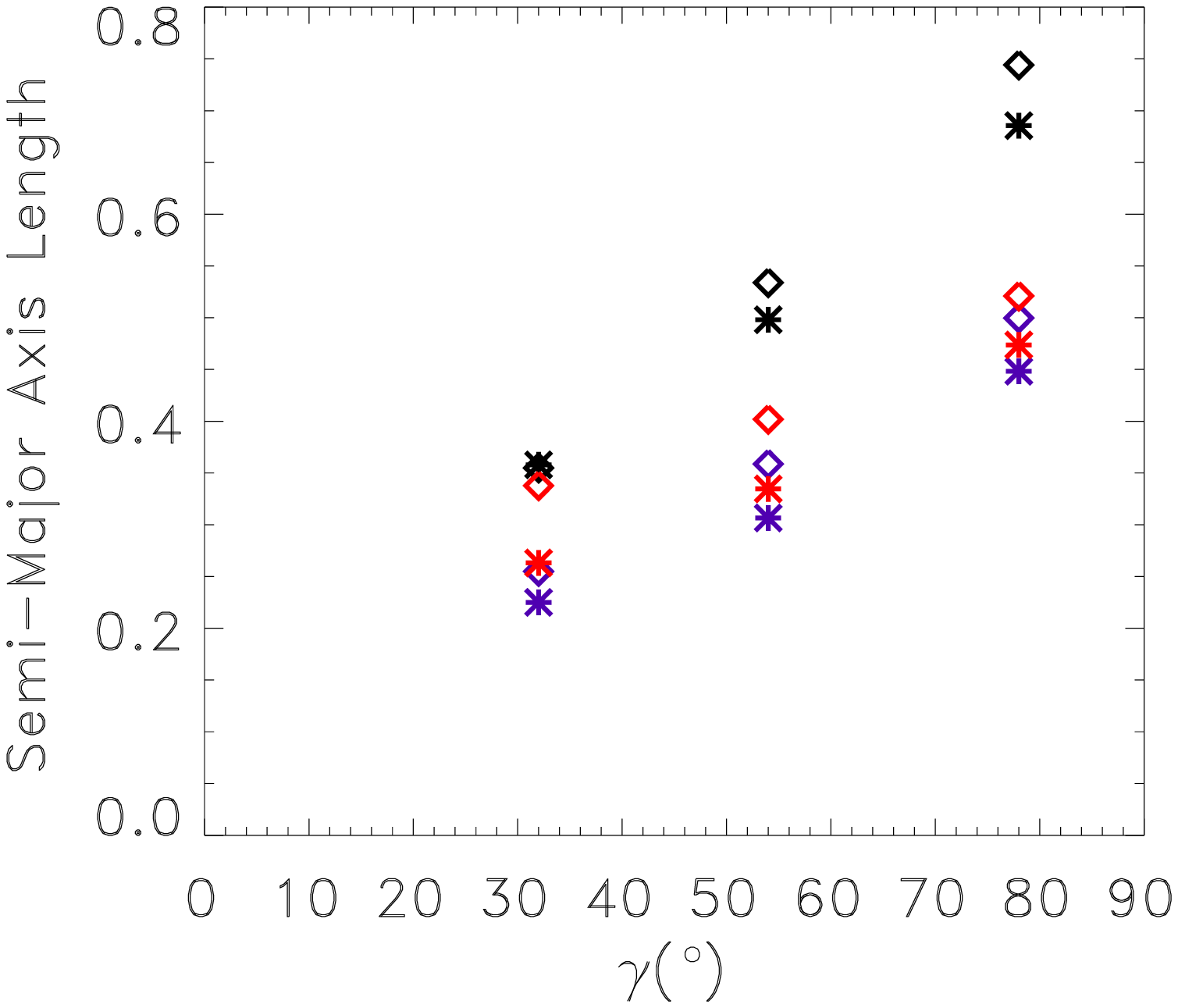}
\caption{Properties of the ellipses; deviation angle, $\delta$ (top left); the 
ellipse eccentricity  for both sunspots at all frequencies against the three 
average magnetic field inclinations (top right); the phase difference between 
the least-squares-fit correlation phase at $\theta_p=-60^\circ$   and at  
$\theta_p=+60^\circ$ ( $\delta \phi_{( \theta_p=-60^\circ) } -  \delta \phi_{( 
\theta_p=+60^\circ) }$ ) (bottom left);  the length of the semi-major axis of 
each ellipse in each region for both sunspots at all frequencies against the 
three average magnetic field inclinations (bottom right). AR9026 is represented 
by the asterisk, and AR9057 by the diamond. 5 mHz is black, 4 mHz is purple and 
3 mHz is red.}
\label{ellp}
\end{center}
\end{figure}

We define a deviation angle 
as the angle between the magnetic field vector and surface
velocity ellipse semi-major axis ($\delta$ = $\gamma - \beta$, where $\beta$ is the inclination of the semi-major axis from vertical).
We show the variation of
the deviation angle, length of the semi-major axis,
and eccentricity with magnetic field strength and/or inclination 
in Figure~\ref{ellp}.
Also shown is the phase difference between the fits at 
$\theta_p=-60^\circ$ and $\theta_p=+60^\circ$. 
This quantity is generally inversely correlated to the ellipse
eccentricity, but is a more 
direct measure of the total variation of observed phase shift
for a specific penumbral region \cite{SBCL05}.
In Figure~\ref{ellp}, data from 
AR9026 is represented by an asterisk, AR9057 by a diamond and the frequencies 
are color-coded as following :- 5 mHz is black, 4 mHz is purple and 3 mHz is 
red.

In general, the trends shown in Figure~\ref{ellp}, namely an increase
in the deviation angle, eccentricity, and semi-major axis length,
and a decrease in the phase variation, with increasing inclination
are observed in both sunspots and at all frequencies. There are
some deviations from this. For example, at low inclinations, it is
observed that the eccenctricity (phase variation) decreases (increases)
with frequency. Note that the semi-major axis length is an indication
of the total wave amplitude in the magnetic region. The trend observed
in the lower-left panel of Figure~\ref{ellp} is consistent with 
a reduction in wave amplitude related to the field strength.

\section{Local Egression Control Correlation}\label{eg}
In this section we 
examine the phase of the local egression control correlation, 
\begin{equation}
C_{+}(\mathbf{r},\nu) = \langle \hat{H}_{+} 
(\mathbf{r},\nu)\hat{\psi}^{*}(\mathbf{r},\nu ) \rangle_{\Delta \nu} = |C_+|  
\rm{e}^{-i \ \delta \phi_{+}}.
\label{correl2}
\end{equation}
Since the egression is simply the time reverse of the 
ingression, we might expect 
to see a reversal of the phase change compared to the ingression phases.

\begin{figure}[p]
\begin{center}
\includegraphics[width=0.9\textwidth]{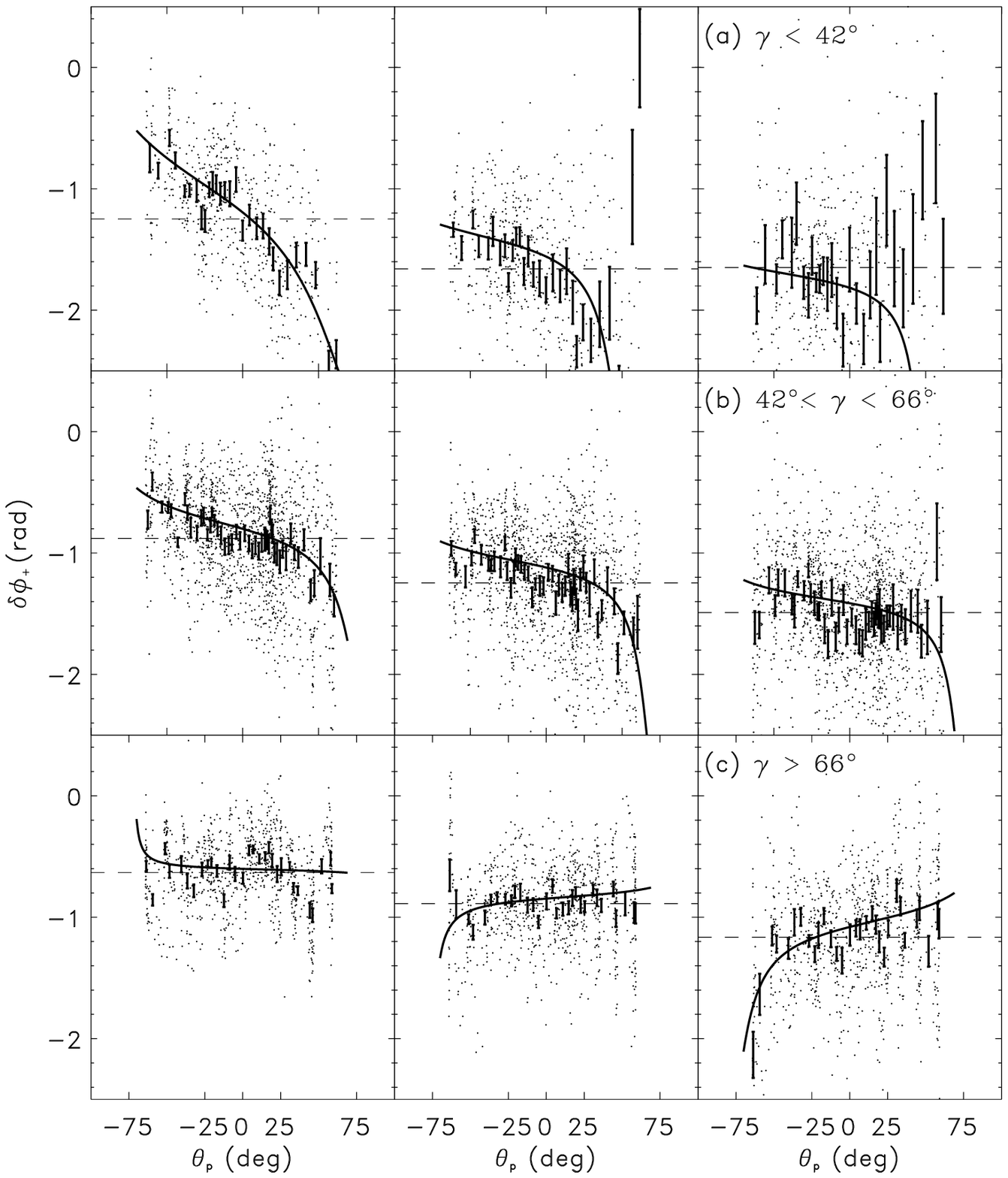} 
\vspace{1cm}
\caption{The 3 mHz (left column), 4 mHz (middle column) and 5 mHz (right 
column) egression correlation phase ($\delta \phi_{+}$) vs. $\theta_p$ within 
the penumbra of sunspot AR9026 for different values of magnetic field 
inclination as indicated.  The three rows represent different portions of the 
penumbra as shown in Figure~\ref{g_vs_b}(a). The top panel row (a) shows 
$\gamma < 42^\circ$, where the mean field strength is $\langle \mathbf{B} 
\rangle = 1900$ G, the middle row (b) shows $42^\circ < \gamma < 66^\circ$, 
where $\langle \mathbf{B} \rangle = 1400$ G, and the bottom row (c) shows  
$\gamma > 66^\circ$, where $\langle \mathbf{B} \rangle = 600$ G. The horizontal 
dashed lines indicate the mean value of $\delta \phi_{+}$ for each panel. }
\label{hephase1}
\end{center}
\end{figure}

\begin{figure}[p]
\begin{center}
\includegraphics[width=0.9\textwidth]{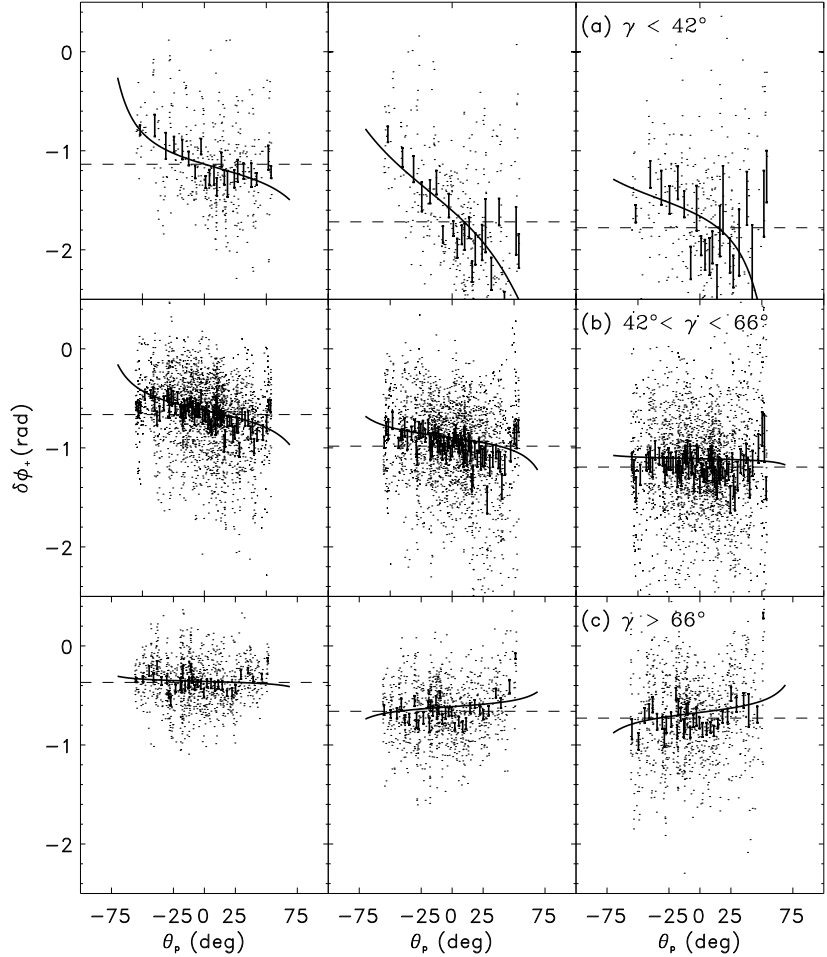}
\vspace{1cm}
\caption{The 3 mHz (left column), 4 mHz (middle column) and 5 mHz (left column) 
egression correlation phase ($\delta \phi_{+}$) vs. $\theta_p$ within the 
penumbra of sunspot AR9057 for different values of magnetic field strength as 
indicated.   The three different panels represent different portions of the 
penumbra, similar to Figure~\ref{g_vs_b}(b). The top panel (a) shows $\gamma < 42^\circ$ where the mean field strength is $\langle B \rangle = 1700$ G the middle panel (b) 
shows $42^\circ < \gamma < 66^\circ$, and $\langle B \rangle = 1000$ G and the 
bottom panel (c) shows $\gamma > 66^\circ$ and $\langle B \rangle = 600$ G. The horizontal dashed lines indicate the mean value of $\delta 
\phi_{+}$ for each panel. }
\label{hephase2}
\end{center}
\end{figure}

\begin{figure}[p]
\begin{center}
\includegraphics[width=0.9\textwidth]{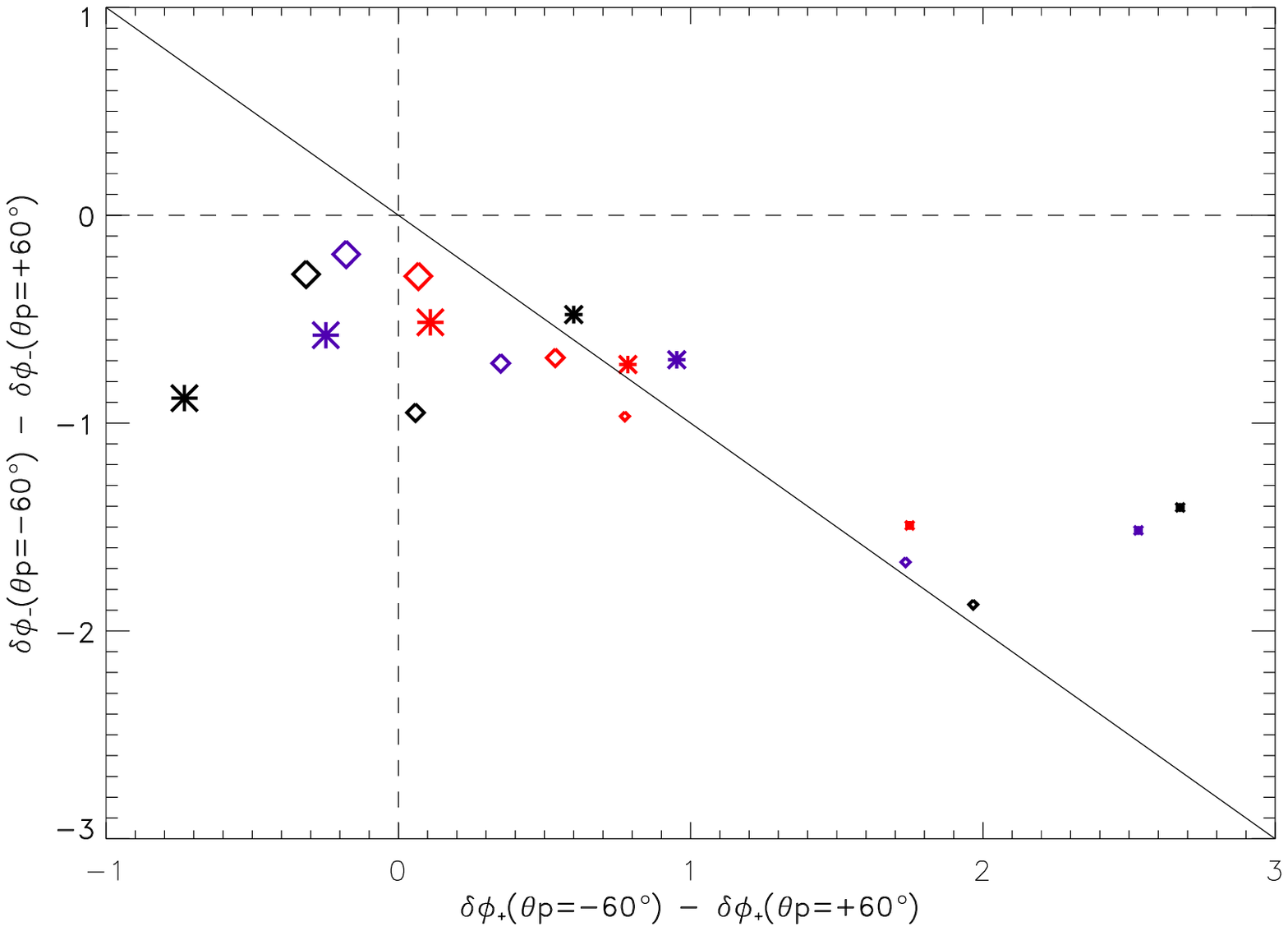}
\vspace{1cm}
\caption{The phase difference between the least-squares-fit correlation phase 
at $\theta_p=-60^\circ$   and at  $\theta_p=+60^\circ$ of the ingression 
plotted against that of the egression. The solid line is a line of slope $-1$. 
AR9026 is represented by an asterisk, AR9057 by a diamond and the 
colors indicate the following frequencies: 5 mHz (black), 4 mHz (purple) and 
3 mHz (red). The largest symbols represent $\gamma > 66^\circ$ and the 
smallest symbols $\gamma < 42^\circ$. }
\label{dphase}
\end{center}
\end{figure}

Using all the days' of data, the egression correlation phase is plotted against 
$\theta_p$ in Figures~\ref{hephase1} and \ref{hephase2}, along with
the fits for the phase variation due to elliptical motion.  We see similar  
trends in both sunspots, AR9026 and AR9787. 
Figure~\ref{dphase} shows the phase 
difference between the least-squares-fit correlation phase at 
$\theta_p=-60^\circ$   and at  $\theta_p=+60^\circ$ of the ingression plotted 
against that of the egression. The colors and symbols represent the same as 
before:  AR9026 is represented by an asterisk, AR9057 by a diamond and the 
frequencies are 5 mHz (black), 4 mHz (purple) and 3 mHz (red) in addition the 
size of the symbols represent the average inclination from vertical. The solid 
line has a slope of $-1$.  There is a reverse 
behaviour of the ingression, compared to the previous egression
results, present for all frequencies at most magnetic field 
inclinations.  The exception is when the field is highly inclined, where 
we observe a trend in the same sense as the ingression. 
This is unexpected 
and warrants further study.

\section{Discussion}

Mode Conversion predicts (among other things \cite{C07}) that when the attack 
angle is small most of the observable energy will be in the slow acoustic mode, 
and when the attack angle is large most of the observable energy will be in the 
fast magnetic mode. In this case we will be seeing the line-of-sight effect of 
a combination of waves coming from all directions impinging on magnetic field 
with a particular orientation. The observations by MDI consist of line-of-sight Doppler signatures 
of the surface motion which are presumably caused mainly by pressure 
perturbations. This means that we would expect to be observing only the slow 
acoustic mode, however it is possible that we are observing a combination of 
the acoustic and magnetic modes.

The dependence of the phase shift of the observed ingression correlations with azimuthal angle 
around sunspot penumbrae, as viewed from different observational vantages, shows that
the incoming phase shifts must be (at least partly) photospheric in origin and are
influenced by the presence of inclined magnetic fields \cite{SBCL05}. 
Analysis of the variation of both the amplitude and phase of the surface velocities 
provides an opportunity to characterize the magnetically influenced acoustic signature
as an ellipse with properties determined by the magnetic fields.  
We found that the ellipses are either nearly vertical (for weaker, more inclined fields) or 
generally directed \emph{away} from the magnetic field direction (for stronger,
more vertical fields).  Largely consistent results for two  active regions,  AR9026 and AR9057, are found.
Some properties of the surface ellipses, e.g. their inclinations, are different for the two sunspots.
Some of this variation may be due to  differences in the field properties. For
example, the field in AR9057 is on 
average $\sim 15\%$ weaker than AR9026. 

Fits of the elliptical motion in Figures~\ref{ellipse9026} and 
\ref{ellipse9057} depend critically on the correlation modulus which is prone 
to systematic uncertainties. But the trend is that a stronger, less 
inclined magnetic field produces elliptical motion with smaller amplitude,
eccentricity, and deviation angle, and a larger inclination from vertical.
These trends exist for both spots and, largely, at all frequencies.
The  shorter 
semi-major axis at strong magnetic field strengths is consistent with previous 
knowledge of surface acoustic amplitude suppression in magnetic fields. It is 
curious to note, however, that at 4 mHz the amplitude is consistently smaller 
than even 3 mHz.

These are the first results to estimate the behaviour of the surface velocity 
ellipse at the photosphere within sunspots.  
The results do not immediately suggest an observation of the slow wave as shown 
by \inlinecite{C05} or \inlinecite{SC06}, but are consistent with their 
expectations of the behaviour of slow waves at this height in the atmosphere.  
Mode conversion theory states the alignment is dependent upon $a^2/c^2$  which 
at the observational heights of $\sim 200$km of the atmosphere may not be large 
enough to invoke alignment.  Since the ray analysis is somewhat unrealistic we 
would expect to observe a combination of fast and slow waves at the surface, 
which will contribute to a clouded view of the surface velocities. 
\inlinecite{RSWS07} are currently exploring the possibility that these apparent 
surface effects are due to the changes in the radiative transfer within active 
regions and the formation height of the observational Ni 678 nm line. 
This explanation requires an absorption mechanism, or else some other means
of producing a difference between the amplitudes of upward and downward
propagating waves. Thus mode conversion may still be important in this proposed
mechanism. A test of the mechanism proposed by Rajaguru et al. (2007) would be to repeat
the observations performed here in a magnetically insensitive line, where the
proposed radiative transfer effects would not be present. In terms of mode conversion, it is 
suggested that the main effect occurs along the bright radial filaments of the 
interlocking comb structure as presented in the penumbral models of 
\inlinecite{WTBT04}. However,  observational helioseismic spatial resolution 
cannot currently resolve this.

This is also the first time that the variation of the phase of the local 
egression correlation has been analysed in the penumbra. It is curious that the 
egression correlation shows a reverse dependence when the magnetic field is 
weak and highly inclined.  This is evidence of a reverse ingression dependence 
on the line-of-sight, but further investigation is required to understand the 
behavior at high frequencies in the weaker, more inclined fields.

\vspace{2cm}

This work was supported in part by the \textit{European Helio- and Asteroseismology Network} (HELAS).



\bibliographystyle{spr-mp-sola}

\bibliography{sola_bib}  

\begin{thebibliography}{27}
\ifx \bisbn   \undefined \def \bisbn  #1{ISBN #1}   \fi
\ifx \binits  \undefined \def \binits#1{#1} \fi
\ifx \bauthor  \undefined \def \bauthor#1{#1} \fi
\ifx \batitle  \undefined \def \batitle#1{#1} \fi
\ifx \bjtitle  \undefined \def \bjtitle#1{#1} \fi
\ifx \bvolume  \undefined \def \bvolume#1{#1} \fi
\ifx \byear  \undefined \def \byear#1{#1} \fi
\ifx \bissue  \undefined \def \bissue#1{#1} \fi
\ifx \bfpage  \undefined \def \bfpage#1{#1} \fi
\ifx \blpage  \undefined \def \blpage #1{#1} \fi
\ifx \burl  \undefined \def \burl#1{#1} \fi
\ifx \binterref  \undefined \def \binterref#1{#1} \fi
\ifx \betal  \undefined \def \betal#1{#1} \fi
\ifx \binstitute  \undefined \def \binstitute#1{#1} \fi
\ifx \bctitle  \undefined \def \bctitle#1{#1} \fi
\ifx \beditor  \undefined \def \beditor#1{#1} \fi
\ifx \bpublisher  \undefined \def \bpublisher#1{#1} \fi
\ifx \bbtitle  \undefined \def \bbtitle#1{#1} \fi
\ifx \bedition  \undefined \def \bedition#1{#1} \fi
\ifx \bseriesno  \undefined \def \bseriesno#1{#1} \fi
\ifx \blocation  \undefined \def \blocation#1{#1} \fi
\ifx \bsertitle  \undefined \def \bsertitle#1{#1} \fi
\ifx \bsnm \undefined \def \bsnm#1{#1} \fi
\ifx \bsuffix \undefined \def \bsuffix#1{#1} \fi
\ifx \bparticle \undefined \def \bparticle#1{#1} \fi
\ifx \barticle \undefined \def \barticle#1{#1} \fi
\ifx \botherref \undefined \def \botherref #1{#1} \fi
\ifx \url \undefined \def \url#1{\textsf{#1}} \fi
\ifx \bchapter \undefined \def \bchapter#1{#1} \fi
\ifx \bbook \undefined \def \bbook#1{#1} \fi
\ifx \bcomment \undefined \def \bcomment#1{#1} \fi
\ifx \oauthor \undefined \def \oauthor#1{#1} \fi
\ifx \citeauthoryear \undefined \def \citeauthoryear#1{#1} \fi
\def \endbibitem {}

\bibitem[\protect\citeauthoryear{{Basu}, {Antia}, and {Bogart}}{2004}]{BAB04}
\begin{barticle}
\bauthor{\bsnm{{Basu}},~\binits{S.}}, \bauthor{\bsnm{{Antia}},~\binits{H.M.}},
  \bauthor{\bsnm{{Bogart}},~\binits{R.S.}}:
\byear{2004}, \batitle{{Ring-Diagram Analysis of the Structure of Solar Active
  Regions}}. \textit{\bjtitle{\apj}} \textbf{\bvolume{610}}, 1157.
\end{barticle}
\endbibitem

\bibitem[\protect\citeauthoryear{{Braun}}{1995}]{B95}
\begin{barticle}
\bauthor{\bsnm{{Braun}},~\binits{D.C.}}:
\byear{1995}, \batitle{{Scattering of p-Modes by Sunspots. I. Observations}}.
  \textit{\bjtitle{\apj}} \textbf{\bvolume{451}}, 859.
\end{barticle}
\endbibitem

\bibitem[\protect\citeauthoryear{{Braun}}{1997}]{B97}
\begin{barticle}
\bauthor{\bsnm{{Braun}},~\binits{D.C.}}:
\byear{1997}, \batitle{{Time-Distance Sunspot Seismology with GONG Data}}.
  \textit{\bjtitle{\apj}} \textbf{\bvolume{487}}, 447.
\end{barticle}
\endbibitem

\bibitem[\protect\citeauthoryear{{Braun} and {Lindsey}}{2000}]{BL00}
\begin{barticle}
\bauthor{\bsnm{{Braun}},~\binits{D.C.}},
  \bauthor{\bsnm{{Lindsey}},~\binits{C.}}:
\byear{2000}, \batitle{{Phase-sensitive Holography of Solar Activity}}.
  \textit{\bjtitle{\solphys}} \textbf{\bvolume{192}}, 307.
\end{barticle}
\endbibitem

\bibitem[\protect\citeauthoryear{{Cally}}{2005}]{C05}
\begin{barticle}
\bauthor{\bsnm{{Cally}},~\binits{P.S.}}:
\byear{2005}, \batitle{{Local magnetohelioseismology of active regions}}.
  \textit{\bjtitle{\mnras}} \textbf{\bvolume{358}}, 353.
\end{barticle}
\endbibitem

\bibitem[\protect\citeauthoryear{{Cally}}{2007}]{C07}
\begin{barticle}
\bauthor{\bsnm{{Cally}},~\binits{P.S.}}:
\byear{2007}, \batitle{{What to look for in the seismology of solar active
  regions}}. \textit{\bjtitle{Astronomische Nachrichten}}
  \textbf{\bvolume{328}}, 286.
\end{barticle}
\endbibitem

\bibitem[\protect\citeauthoryear{{Cally}, {Crouch}, and {Braun}}{2003}]{CCB03}
\begin{barticle}
\bauthor{\bsnm{{Cally}},~\binits{P.S.}},
  \bauthor{\bsnm{{Crouch}},~\binits{A.D.}},
  \bauthor{\bsnm{{Braun}},~\binits{D.C.}}:
\byear{2003}, \batitle{{Probing sunspot magnetic fields with p-mode absorption
  and phase shift data}}. \textit{\bjtitle{\mnras}} \textbf{\bvolume{346}},
  381.
\end{barticle}
\endbibitem

\bibitem[\protect\citeauthoryear{{Couvidat}, {Birch}, and
  {Kosovichev}}{2006}]{CBK06}
\begin{barticle}
\bauthor{\bsnm{{Couvidat}},~\binits{S.}},
  \bauthor{\bsnm{{Birch}},~\binits{A.C.}},
  \bauthor{\bsnm{{Kosovichev}},~\binits{A.G.}}:
\byear{2006}, \batitle{{Three-dimensional Inversion of Sound Speed below a
  Sunspot in the Born Approximation}}. \textit{\bjtitle{\apj}}
  \textbf{\bvolume{640}}, 516.
\end{barticle}
\endbibitem

\bibitem[\protect\citeauthoryear{{Couvidat} and {Rajaguru}}{2007}]{CR07}
\begin{barticle}
\bauthor{\bsnm{{Couvidat}},~\binits{S.}},
  \bauthor{\bsnm{{Rajaguru}},~\binits{S.P.}}:
\byear{2007}, \batitle{{Contamination by Surface Effects of Time-Distance
  Helioseismic Inversions for Sound Speed beneath Sunspots}}.
  \textit{\bjtitle{\apj}} \textbf{\bvolume{661}}, 558.
\end{barticle}
\endbibitem

\bibitem[\protect\citeauthoryear{{Crouch} and {Cally}}{2003}]{CC03}
\begin{barticle}
\bauthor{\bsnm{{Crouch}},~\binits{A.D.}},
  \bauthor{\bsnm{{Cally}},~\binits{P.S.}}:
\byear{2003}, \batitle{{Mode Conversion of Solar p Modes in non-Vertical
  Magnetic Fields - i. two-Dimensional Model}}. \textit{\bjtitle{\solphys}}
  \textbf{\bvolume{214}}, \bfpage{201}\,--\,\blpage{226}.
\end{barticle}
\endbibitem

\bibitem[\protect\citeauthoryear{{Duvall} \textit{et~al.}}{1993}]{DJHP93}
\begin{barticle}
\bauthor{\bsnm{{Duvall}},~\binits{T.L.}\bsuffix{Jr.}},
  \bauthor{\bsnm{{Jefferies}},~\binits{S.M.}},
  \bauthor{\bsnm{{Harvey}},~\binits{J.W.}},
  \bauthor{\bsnm{{Pomerantz}},~\binits{M.A.}}:
\byear{1993}, \batitle{{Time-distance helioseismology}}.
  \textit{\bjtitle{\nat}} \textbf{\bvolume{362}}, 430.
\end{barticle}
\endbibitem

\bibitem[\protect\citeauthoryear{{Hanasoge} \textit{et~al.}}{2007}]{HCRB07}
\begin{botherref}
\oauthor{\bsnm{{Hanasoge}},~\binits{S.M.}},
  \oauthor{\bsnm{{Couvidat}},~\binits{S.}},
  \oauthor{\bsnm{{Rajaguru}},~\binits{S.P.}},
  \oauthor{\bsnm{{Birch}},~\binits{A.C.}}:
2007, {Impact of Locally Suppressed Wave sources on helioseismic travel times}.
  \textit{ArXiv e-prints} \textbf{707}.
\end{botherref}
\endbibitem

\bibitem[\protect\citeauthoryear{{Korzennik}}{2006}]{K06}
\begin{botherref}
\oauthor{\bsnm{{Korzennik}},~\binits{S.G.}}:
2006, {The cookie cutter test for time-distance tomography of active regions}.
  In: \textit{Proceedings of SOHO 18/GONG 2006/HELAS I, Beyond the spherical
  Sun}, \textit{ESA Special Publication}, \textbf{624}.
\end{botherref}
\endbibitem

\bibitem[\protect\citeauthoryear{{Kosovichev}, {Duvall}, and
  {Scherrer}}{2000}]{KDS00}
\begin{barticle}
\bauthor{\bsnm{{Kosovichev}},~\binits{A.G.}},
  \bauthor{\bsnm{{Duvall}},~\binits{T.L..J.}},
  \bauthor{\bsnm{{Scherrer}},~\binits{P.H.}}:
\byear{2000}, \batitle{{Time-Distance Inversion Methods and Results - (Invited
  Review)}}. \textit{\bjtitle{\solphys}} \textbf{\bvolume{192}}, 159.
\end{barticle}
\endbibitem

\bibitem[\protect\citeauthoryear{{Lindsey} and {Braun}}{2000}]{LB00}
\begin{barticle}
\bauthor{\bsnm{{Lindsey}},~\binits{C.}},
  \bauthor{\bsnm{{Braun}},~\binits{D.C.}}:
\byear{2000}, \batitle{{Basic Principles of Solar Acoustic Holography -
  (Invited Review)}}. \textit{\bjtitle{\solphys}} \textbf{\bvolume{192}},
  \bfpage{261}\,--\,\blpage{284}.
\end{barticle}
\endbibitem

\bibitem[\protect\citeauthoryear{{Lindsey} and {Braun}}{2005}]{LB05i}
\begin{barticle}
\bauthor{\bsnm{{Lindsey}},~\binits{C.}},
  \bauthor{\bsnm{{Braun}},~\binits{D.C.}}:
\byear{2005}, \batitle{{The Acoustic Showerglass. I. Seismic Diagnostics of
  Photospheric Magnetic Fields}}. \textit{\bjtitle{\apj}}
  \textbf{\bvolume{620}}, 1107.
\end{barticle}
\endbibitem

\bibitem[\protect\citeauthoryear{{Lindsey} and {Braun}}{2005}]{LB05ii}
\begin{barticle}
\bauthor{\bsnm{{Lindsey}},~\binits{C.}},
  \bauthor{\bsnm{{Braun}},~\binits{D.C.}}:
\byear{2005}, \batitle{{The Acoustic Showerglass. II. Imaging Active Region
  Subphotospheres}}. \textit{\bjtitle{\apj}} \textbf{\bvolume{620}}, 1118.
\end{barticle}
\endbibitem

\bibitem[\protect\citeauthoryear{{Mickey} \textit{et~al.}}{1996}]{IVM96}
\begin{barticle}
\bauthor{\bsnm{{Mickey}},~\binits{D.L.}},
  \bauthor{\bsnm{{Canfield}},~\binits{R.C.}},
  \bauthor{\bsnm{{Labonte}},~\binits{B.J.}},
  \bauthor{\bsnm{{Leka}},~\binits{K.D.}},
  \bauthor{\bsnm{{Waterson}},~\binits{M.F.}},
  \bauthor{\bsnm{{Weber}},~\binits{H.M.}}:
\byear{1996}, \batitle{{The Imaging Vector Magnetograph at Haleakala}}.
  \textit{\bjtitle{\solphys}} \textbf{\bvolume{168}}, 229.
\end{barticle}
\endbibitem

\bibitem[\protect\citeauthoryear{{Rajaguru} \textit{et~al.}}{2006}]{RBDTZ06}
\begin{barticle}
\bauthor{\bsnm{{Rajaguru}},~\binits{S.P.}},
  \bauthor{\bsnm{{Birch}},~\binits{A.C.}},
  \bauthor{\bsnm{{Duvall}},~\binits{T.L.}\bsuffix{Jr.}},
  \bauthor{\bsnm{{Thompson}},~\binits{M.J.}},
  \bauthor{\bsnm{{Zhao}},~\binits{J.}}:
\byear{2006}, \batitle{{Sensitivity of Time-Distance Helioseismic Measurements
  to Spatial Variation of Oscillation Amplitudes. I. Observations and a
  Numerical Model}}. \textit{\bjtitle{\apj}} \textbf{\bvolume{646}}, 543.
\end{barticle}
\endbibitem

\bibitem[\protect\citeauthoryear{{Rajaguru} \textit{et~al.}}{2007}]{RSWS07}
\begin{barticle}
\bauthor{\bsnm{{Rajaguru}},~\binits{S.P.}},
  \bauthor{\bsnm{{Sankarasubramanian}},~\binits{K.}},
  \bauthor{\bsnm{{Wachter}},~\binits{R.}},
  \bauthor{\bsnm{{Scherrer}},~\binits{P.H.}}:
\byear{2007}, \batitle{{Radiative Transfer Effects on Doppler Measurements as
  Sources of Surface Effects in Sunspot Seismology}}. \textit{\bjtitle{\apjl}}
  \textbf{\bvolume{654}}, L175.
\end{barticle}
\endbibitem

\bibitem[\protect\citeauthoryear{{Scherrer} \textit{et~al.}}{1995}]{MDI95}
\begin{barticle}
\bauthor{\bsnm{{Scherrer}},~\binits{P.H.}},
  \bauthor{\bsnm{{Bogart}},~\binits{R.S.}},
  \bauthor{\bsnm{{Bush}},~\binits{R.I.}},
  \bauthor{\bsnm{{Hoeksema}},~\binits{J.T.}},
  \bauthor{\bsnm{{Kosovichev}},~\binits{A.G.}},
  \bauthor{\bsnm{{Schou}},~\binits{J.}},
  \bauthor{\bsnm{{Rosenberg}},~\binits{W.}},
  \bauthor{\bsnm{{Springer}},~\binits{L.}},
  \bauthor{\bsnm{{Tarbell}},~\binits{T.D.}},
  \bauthor{\bsnm{{Title}},~\binits{A.}},
  \bauthor{\bsnm{{Wolfson}},~\binits{C.J.}},
  \bauthor{\bsnm{{Zayer}},~\binits{I.}}, \bauthor{\bsnm{{MDI Engineering
  Team}},~}:
\byear{1995}, \batitle{{The Solar Oscillations Investigation - Michelson
  Doppler Imager}}. \textit{\bjtitle{\solphys}} \textbf{\bvolume{162}}, 129.
\end{barticle}
\endbibitem

\bibitem[\protect\citeauthoryear{{Schunker}, {Braun}, and
  {Cally}}{2007}]{SBC07}
\begin{barticle}
\bauthor{\bsnm{{Schunker}},~\binits{H.}},
  \bauthor{\bsnm{{Braun}},~\binits{D.C.}},
  \bauthor{\bsnm{{Cally}},~\binits{P.S.}}:
\byear{2007}, \batitle{{Surface magnetic field effects in local
  helioseismology}}. \textit{\bjtitle{Astronomische Nachrichten}}
  \textbf{\bvolume{328}}, 292.
\end{barticle}
\endbibitem

\bibitem[\protect\citeauthoryear{{Schunker} \textit{et~al.}}{2005}]{SBCL05}
\begin{barticle}
\bauthor{\bsnm{{Schunker}},~\binits{H.}},
  \bauthor{\bsnm{{Braun}},~\binits{D.C.}},
  \bauthor{\bsnm{{Cally}},~\binits{P.S.}},
  \bauthor{\bsnm{{Lindsey}},~\binits{C.}}:
\byear{2005}, \batitle{{The Local Helioseismology of Inclined Magnetic Fields
  and the Showerglass Effect}}. \textit{\bjtitle{\apjl}}
  \textbf{\bvolume{621}}, L149.
\end{barticle}
\endbibitem

\bibitem[\protect\citeauthoryear{{Schunker} and {Cally}}{2006}]{SC06}
\begin{barticle}
\bauthor{\bsnm{{Schunker}},~\binits{H.}},
  \bauthor{\bsnm{{Cally}},~\binits{P.S.}}:
\byear{2006}, \batitle{{Magnetic field inclination and atmospheric oscillations
  above solar active regions}}. \textit{\bjtitle{\mnras}}
  \textbf{\bvolume{372}}, 551.
\end{barticle}
\endbibitem

\bibitem[\protect\citeauthoryear{{Weiss} \textit{et~al.}}{2004}]{WTBT04}
\begin{barticle}
\bauthor{\bsnm{{Weiss}},~\binits{N.O.}},
  \bauthor{\bsnm{{Thomas}},~\binits{J.H.}},
  \bauthor{\bsnm{{Brummell}},~\binits{N.H.}},
  \bauthor{\bsnm{{Tobias}},~\binits{S.M.}}:
\byear{2004}, \batitle{{The Origin of Penumbral Structure in Sunspots: Downward
  Pumping of Magnetic Flux}}. \textit{\bjtitle{\apj}} \textbf{\bvolume{600}},
  1073.
\end{barticle}
\endbibitem

\bibitem[\protect\citeauthoryear{{Zhao} and {Kosovichev}}{2003}]{ZK03}
\begin{barticle}
\bauthor{\bsnm{{Zhao}},~\binits{J.}},
  \bauthor{\bsnm{{Kosovichev}},~\binits{A.G.}}:
\byear{2003}, \batitle{{Helioseismic Observation of the Structure and Dynamics
  of a Rotating Sunspot Beneath the Solar Surface}}. \textit{\bjtitle{\apj}}
  \textbf{\bvolume{591}}, 446.
\end{barticle}
\endbibitem

\bibitem[\protect\citeauthoryear{{Zhao} and {Kosovichev}}{2006}]{ZK06}
\begin{barticle}
\bauthor{\bsnm{{Zhao}},~\binits{J.}},
  \bauthor{\bsnm{{Kosovichev}},~\binits{A.G.}}:
\byear{2006}, \batitle{{Surface Magnetism Effects in Time-Distance
  Helioseismology}}. \textit{\bjtitle{\apj}} \textbf{\bvolume{643}}, 1317.
\end{barticle}
\endbibitem

\end{thebibliography}

\IfFileExists{\jobname.bbl}{} {\typeout{}
\typeout{****************************************************}
\typeout{****************************************************}
\typeout{** Please run "bibtex \jobname" to obtain} \typeout{**
the bibliography and then re-run LaTeX} \typeout{** twice to fix
the references !}
\typeout{****************************************************}
\typeout{****************************************************}
\typeout{}}

\clearpage

\end{article} 
\end{document}